\documentclass[12pt]{article}
\usepackage{axodraw}

\newcommand{\beq}{\begin{equation}}
\newcommand{\eeq}{\end{equation}}
\newcommand{\beqa}{\begin{eqnarray}}
\newcommand{\eeqa}{\end{eqnarray}}
\newcommand{\beqar}{\begin{eqnarray*}}
\newcommand{\eeqar}{\end{eqnarray*}}

\newcommand{\al}{\alpha}
\newcommand{\be}{\beta}

\def\non          {\nonumber}

\def\Tr           {\mbox{\rm Tr}\,}

\def\cd           {{\cdot}}
\def\ran          {\rangle}
\def\lan          {\langle}
\def\fsH    {H\!\!\!\!/\,}
\def\fsC    {C\!\!\!\!/\,}

\newcommand{\eps}{\epsilon}
\newcommand{\ga}{\gamma}

\newcommand{\inn}{\!\cdot\!}

\newcommand{\lam}{\lambda}

\newcommand{\z}{\zeta}

\newcommand{\labell}[1]{\label{#1}} 
\newcommand{\reef}[1]{(\ref{#1})}
\newcommand\prt{\partial}

\newcommand\cD{{\cal D}}

\begin{document}
\baselineskip 17pt%
\begin{titlepage}
\vspace*{.8mm}%
\vspace*{8mm}%

\center{ {\bf \Large  All order $\alpha'$ higher derivative corrections to non-BPS branes of type IIB Super string theory

}}\vspace*{1mm} \centerline{{\Large {\bf  }}}
\vspace*{3mm}
\begin{center}
{Ehsan Hatefi }$\footnote{E-mail:ehatefi@ictp.it}$

\vspace*{0.6cm}{ {\it
International Centre for Theoretical Physics\\
 Strada Costiera 11, Trieste, Italy
  \\

}}

\vspace*{.2cm}
\end{center}
\begin{center}{\bf Abstract}\end{center}
\begin{quote}

By dealing with the evaluation of string theory correlators of $<V_C V_T V_{\bar\Psi} V_\Psi>$, the complete and closed form of  the amplitude of two fermion fields, one tachyon and one closed string Ramond-Ramond field in type IIB superstring theory is found. Specifically by comparing infinite tachyon poles in  field theory amplitude with infinite tachyon poles of the S-matrix of string amplitude (for $p+1=n$ case),  all the infinite higher derivative  corrections of two tachyons and two fermions (in type IIB) to all orders of $\alpha'$  have been discovered. Using these new couplings, we are able to produce  infinite $t'+s'+u$-channel tachyon poles of string theory in field theory. Due to internal degrees of freedom of fermions and tachyon (Chan-Paton factors) we comment that, neither there should be  single $s,t-$channel  fermion (tachyon  pole)  nor their infinite poles.

Due to internal CP factor we also discover that there is no coupling between two closed string Ramond-Ramond (RR) field and  one tachyon in type II super string theory.

Taking into account the fact that the kinetic terms of fermions, gauge, scalar fields and tachyons do not obtain any higher derivative corrections, and due to their CP factors, string theory amplitude dictates us there should not be any double poles in the amplitude of one RR, two fermions and one tachyon.


\end{quote}
\end{titlepage}
\section{Introduction}

 $D_{p}$-branes must be interpreted just as sources for Ramond-Ramond (closed string C-field) for both BPS and non-BPS branes
 \cite{Polchinski:1995mt,Witten:1995im}. Applying  Ramond-Ramond (RR) couplings various issues  such as  brane within branes  \cite{Douglas:1995bn,Douglas:1997ch} , K-theory in the language of D-branes \cite{Minasian:1997mm,Witten:1998cd}, Myers effect
  \cite{Myers:1999ps,Howe:2006rv}
  and its all order corrections \cite{Hatefi:2012zh} have been well understood.

\vskip .2in

It is also worth trying  to follow  Born-Infeld action  and its generalization which was appeared in \cite{Leigh:1989jq,Tseytlin:1999dj}.

\vskip .1in

By dealing with   unstable branes, one might hope to reveal some of the properties of type IIB (IIA) String theories within some special backgrounds  (more precisely only time-dependent ones). For more explanations we refer to some of the basic references  \cite{Gutperle:2002ai,Sen:2002nu,Sen:2002in,Sen:2002an,Sen:2002vv,
Mukhopadhyay:2002en,Strominger:2002pc,Okuda:2002yd,Lambert:2003zr,Sen:2004nf}.

\vskip .2in

 Having used tachyonic action \cite{Sen:1999md} and in particular, using its correct and complete form (based on S-Matrix calculations, including the internal degrees of freedom of tachyons in different pictures)\cite{Hatefi:2012wj}, one is able to describe most of the known properties of the decay of the unstable $D_{p}$-branes (where  spatial dimension of brane $p$ is even (odd) for IIB (IIA) theories)
just for stable point inside tachyon' s potential.

 \vskip .1in

As argued in detail in  \cite{Hatefi:2012wj} the higher derivative corrections of tachyonic action ( including both non-BPS and D-brane anti D-brane effective actions) are indeed important around their unstable point. In section 2 of  \cite{Hatefi:2012wj} we have made several  aims and motivations for following  unstable branes   nevertheless  let us just  point out some of them once more.

 \vskip .2in

As an instance  for some holographic models with symmetry breaking in QCD one has to use tachyonic action and its higher derivative corrections \cite{Casero:2007ae,Dhar:2008um}. The other motivation is as follows.

\vskip .1in

In order to  deal with inflation in string theory \cite{Dvali:1998pa,Burgess:2001fx,Kachru:2003sx} , one has to make use of the effective action of brane -anti brane system involving its correct higher derivative corrections \cite{Garousi:2007fk,Hatefi:2012cp}. For more explanations  section 2 of \cite {Hatefi:2012wj} is realized. For the other applications on non-BPS branes one might look at \cite{Choudhury:2003vr}.

 \vskip .2in

One way for finding these higher derivative corrections is to apply directly the scattering theory of non-BPS branes. Indeed by applying this method we are able to actually explore new couplings including their corrections and essentially in a very exact manner  their coefficients can be fixed.

\vskip .1in

 However due to S-matrix formalism, around the unstable point of tachyonic DBI action,
 the internal degrees of freedom of tachyons  must be taken into account, namely open string tachyons  in  (-1)-picture do carry $\sigma_2$ and in zero picture  they carry $\sigma_1$ Pauli matrix.

 \vskip .2in

  On the other hand we have constructed and checked the exact conditions for the other part of the effective actions, basically the Wess-Zumino part of both brane anti brane and non-BPS branes
  will give rise all order terms of the resulting amplitudes.

\vskip .1in

  For good reasons we would like to highlight the main works \cite{Kennedy:1999nn,Billo:1999tv,Garousi:2007fk,Garousi:2008ge}.
  Notice that just some of these couplings  can be discovered by making use of BSFT formalism \cite{Kraus:2000nj}.

\vskip .1in

In order to observe some of the applications for all the infinite higher derivative corrections on  BPS branes, $N^3$ entropy of $M5$ branes (regarding Dielectric effect and Black brane entropy growth)  \cite{Hatefi:2012sy} has been suggested. Concerning the applications on M-theory \cite{Hatefi:2012bp,McOrist:2012yc} may be considered.

 \vskip .2in

It is worth trying to talk about super symmetrized version of tachyonic DBI  action. The fermions are also embedded in tachyonic DBI action \cite{Bergshoeff:2000dq}. Having removed tachyons in this action , one may re-derive  the super symmetric version of DBI action   ( for further details  \cite{Aganagic:1996nn,Aganagic:1997zk} should be considered).

 \vskip .1in

 From now on we would like to keep tachyons for our explicit computations. Keeping them fixed  to the action ( the static gauge and special normalization for their kinetic term are employed),  we obtain the Lagrangian as follows

 \vskip .1in

\beqa
L=-T_pV(T)\sqrt{-\det(\eta_{ab}+2\pi\alpha'F_{ab}-2\pi\alpha'\bar\Psi\ga_b\prt_a\Psi+\pi^2\alpha'^2\bar\Psi\ga^{\mu}\prt_a\Psi\bar\Psi\ga_{\mu}\prt_b\Psi + 2\pi\alpha'\prt_aT\prt_bT )}
\nonumber
\eeqa


 \vskip .1in

Remarks on tachyon potential with different approaches have been mentioned (see \cite{Sen:1999mg,DeSmet:2000je,Hatefi:2012wj}), however  it is good to know that the potential which is used to work in the S-matrix computations in super string theory is
\beqa
V(T)=e^{-\pi TT/2}\nonumber\eeqa

 In particular its expansion up to fourth order makes consistent results with scattering amplitude arguments. The other point should be made, is that as tachyon goes to infinity the term inside the effective action  $T^4V(TT)$ tends to zero as we have already expected from  a single unstable brane's condensation.

 \vskip .2in

All the infinite non-Abelian  higher derivative corrections of two fermions two tachyons are not embedded into this action. Here we would like to perform the S-matrix calculations at disk level of one closed string Ramond-Ramond (C-field), one tachyon and two world volume fermions in type IIB super string theory to explore  non-Abelian couplings of two fermions and two tachyons to all orders in $\alpha'$. Obviously by looking for two fermion two tachyon 's amplitude one can fix the needed coefficients in  field theory as we will go through them in detail. An important point has to be clarified as follows:

 \vskip .2in

 By carrying out just this S-matrix $(CT\bar \psi \psi)$, one can precisely fix the coefficients of all higher derivative corrections of  two fermions two tachyons in the world volume of non-BPS branes.

 \vskip .2in

Note that there should not be any coupling between two fermions and one tachyon in the world volume of non-BPS branes as we comment it later on.

 \vskip .2in

Therefore we come to the important fact for which this four point (technically five point) function $(CT\bar \psi \psi)$,
  is the only exception so far which does not include any $s,t,u$ -channel poles.


\section{ Notations and remarks on scattering of non-BPS branes }

Before moving to finding out our amplitude, mentioning some remarks on the scattering of non-BPS branes is inevitable.
Apart from some needed notations, here we want to show that although the amplitude of two world volume fermions and one tachyon in II string theories has non-zero value, however  such a coupling in field theory is not certainly allowed.

\vskip.1in

The reason for this conclusion is that both kinetic terms of fermion fields and tachyons have to carry two fermions and two tachyons.
 We come over to this problem by relabeling the internal degrees of freedom to both left and right hand side of the vertex operator of  Ramond field.

\vskip.2in

It is worth to address some of the works which have been completely dealt with scattering amplitudes of non-BPS \cite{Kennedy:1999nn,Garousi:2008ge,Hatefi:2012wj,Hatefi:2012cp} and BPS branes \cite{Hatefi:2010ik,Stieberger:2009hq,Billo:1998vr,Hatefi:2012ve,Chandia:2003sh,Hatefi:2012rx,Medina:2002nk,Hatefi:2012zh,Barreiro:2005hv,Hatefi:2013eia} at tree level computations.

 \vskip.2in

To achieve our goals, we need to remind the structure of the needed vertex operators \footnote{here we  keep  $\alpha'$ and for the moment do not talk about  internal CP degrees of freedom of the vertex operators }

\beqa
V_{T}^{(0)}(x) &=& \alpha' ik\cd \psi(x)    e^{\alpha' ik\cd X(x)}\nonumber\\
V_{T}^{(-1)}(x) &=&e^{-\phi(x)} e^{\alpha' ik\cd X(x)}\nonumber\\
V_{\bar\Psi}^{(-1/2)}(x)&=&\bar u^Ae^{-\phi(x)/2}S_A(x)\,e^{\alpha'iq.X(x)} \nonumber\\
V_{\Psi}^{(-1/2)}(x)&=&u^Be^{-\phi(x)/2}S_B(x)\,e^{\alpha'iq.X(x)} \nonumber\\
V_{C}^{(-\frac{1}{2},-\frac{1}{2})}(z,\bar{z})&=&(P_{-}\fsH_{(n)}M_p)^{\al\be}e^{-\phi(z)/2}
S_{\al}(z)e^{i\frac{\alpha'}{2}p\cd X(z)}e^{-\phi(\bar{z})/2} S_{\be}(\bar{z})
e^{i\frac{\alpha'}{2}p\cd D \cd X(\bar{z})}
\label{d4Vs}
\eeqa

 The on-shell conditions for all strings, including fermions, Ramond-Ramond and tachyon are
$q^2=p^2=0,k^2=\frac{1}{2 \alpha'}$. Majorana-Weyl ($\bar u^A, u^B$) function  in all ten dimensions have been introduced in \cite{Hatefi:2013eia}. It has already been pointed out that one has to employ charge conjugation  $C^{\alpha\be}$ as well. Let us just emphasize on the structure of RR's field strength with
    $n=1,3,5$ and  $a_n=1$  in type IIB super string theory as follows

\begin{displaymath}
\fsH_{(n)} = \frac{a
_n}{n!}H_{\mu_{1}\ldots\mu_{n}}\ga^{\mu_{1}}\ldots
\ga^{\mu_{n}}
\ ,
\non\end{displaymath}

\vskip 0.1in

A useful trick (doubling trick) in order just to  make use of the holomorphic correlations has been applied. Further notations can be obtained in  \cite{Hatefi:2012wj,Hatefi:2013eia}.

\vskip 0.1in

Notice to the important point that the vertices of two fermion fields are similar to the vertex of  a closed string Ramond-Ramond field so one may expect that the amplitude of two fermions and one tachyon can be derived from the amplitude of one RR  and one tachyon in the world volume of non-BPS branes, however due to their internal Chan-Paton factors we comment that  it is no longer true.

\vskip 0.2in

The amplitude of two fermions and one tachyon  in the world volume of type II string theory without taking into account the internal degrees of freedom (Chan-Paton factors) is given as follows:

\beqa
{\cal A}^{   \bar\psi \psi T} & \sim & \int dx_{1}dx_{2}dx_{3}
  \lan
V_{\bar\psi}^{(-1/2)}{(x_{1})}V_{\psi}^{(-1/2)}{(x_{2})} V_{T}^{(-1)}{(x_{3})}
 \ran,\labell{sstring}\eeqa

Having replaced the mentioned vertex operators in the amplitude, we reach to

\beqa {\cal A}^{\bar\psi, \psi, T}&\sim& \int
 dx_{1}dx_{2}dx_{3}
  x_{12}^{-1/4}(x_{13}x_{23})^{-1/2}|x_{12}|^{\alpha'^2 k_1.k_2}|x_{13}|^{\alpha'^2 k_1.k_3}
 |x_{23}|^{\alpha'^2 k_2.k_3}\nonumber\\&&
\times<:S_{A}(x_1):S_{B}(x_2):>\bar u_1^A u_2^B\label{nop}\eeqa

using
\beqa
 <:S_{A}(x_1):S_{B}(x_2):>= x_{12}^{-5/4}
(C^{-1})_{AB}.\nonumber
\eeqa

and applying the  holomorphic correlators for all the fields

 \beqa
\lan X^{\mu}(z)X^{\nu}(w)\ran & = & -\frac{\alpha'}{2}\eta^{\mu\nu}\log(z-w) \ , \non \\
 \lan \psi^{\mu}(z)\psi^{\nu}(w) \ran & = & -\frac{\alpha'}{2}\eta^{\mu\nu}(z-w)^{-1} \ ,\non \\
\lan\phi(z)\phi(w)\ran & = & -\log(z-w) \ .
\labell{prop}\eeqa

one can explicitly check the $SL(2,R)$ invariance of this amplitude. Gauge fixing as
$(x_1,x_2,x_3)=(0,1,\infty)$  into \reef{nop}, the amplitude should be read as

 \beqa
{\cal A}^{\bar \Psi,\Psi,T} &= &  \bar u_1^A C^{-1}_{AB}u_2^B \left(\Tr(\lam_1\lam_2\lam_3)-\Tr(\lam_1\lam_3\lam_2)\right)\labell{amp2}
\eeqa

which has non-zero value, however  the final result of string theory can not be reproduced in field theory by extracting the kinetic term of the fermion fields ($2\pi\alpha'\bar\psi\ga^{\mu} D_{\mu}\psi$) .

\vskip 0.2in

Also note that the tachyon potential is an even function of tachyon field and also in the DBI effective action both fermion and tachyon fields are even functions, so we come to the fact that this amplitude should have zero result in field theory as well as in string theory side.

The only possibility for not having such a coupling  is to devote  Chan-Paton factors to fermion fields.
 \vskip 0.1in

 The internal degree  of freedom of massless gauge field in (-1)-picture for non-BPS branes  has already been fixed (to be $\sigma_3$) in \cite{Hatefi:2012wj}. Since the amplitude of two fermions and one gauge field has non zero value even for non-BPS branes (as this has been calculated and shown that it has certainly non zero value \cite{jp98}) , we come to the fact  that the internal multiplication of CP factors of  two fermion fields  must carry  $\sigma_3$ in order for getting non zero value for the amplitude of  $A^{\bar\psi^{-1/2}\psi^{-1/2}A^{-1} }$ .

 Hence one can postulate
this internal degree of freedom ($ \sigma_3$) to right hand side of massless Ramond vertex operator and
apply picture changing operator (which carries $ \sigma_3$ CP factor  \cite{DeSmet:2000je} ) on it to get CP factor of left hand side of the Ramond vertex operator which turned out to be an identity matrix (albeit there is an ambiguity about devoting the CP factor of $\sigma_3$ to right hand side or left hand side of Ramond vertex operator, however for our purpose we need not worry about this ambiguity ).

\vskip.2in

On the other hand the $CT$ amplitude   in the world volume of non-BPS branes has certainly non zero value such that by fixing $(x_1= \infty ,z=i,\bar z=-i) $  we obtain the final form of the amplitude of one closed string RR and one tachyon as

\beqa
 A^{CT} &=& (2i)\Tr(P_{-}\fsH_{(n)} M_p)
 \nonumber\eeqa
This result should  be reproduced by taking into account  the following coupling

\beqa
2i\beta'\mu'_p (2\pi\alpha') \int_{\Sigma_{p+1}}C_p\wedge DT
\nonumber\eeqa

Therefore we need to postulate the following CP factors to all vertices in the presence of non-BPS branes

\beqa
V_{T}^{(0)}(x) &=&  \alpha' ik\cd\psi(x) e^{\alpha' ik\cd X(x)}\lam\otimes\sigma_1,
\nonumber\\
V_{T}^{(-1)}(x) &=&e^{-\phi(x)} e^{\alpha' ik\cd X(x)}\lam\otimes\sigma_2\nonumber\\
V_{\bar\Psi}^{(-1/2)}(x)&=&\bar u^Ae^{-\phi(x)/2}S_A(x)\,e^{ \alpha'iq.X(x)}\lam\otimes\sigma_3 \nonumber\\
V_{\Psi}^{(-1/2)}(x)&=&u^Be^{-\phi(x)/2}S_B(x)\,e^{ \alpha'  iq.X(x)}\lam\otimes I \nonumber\\
V_{C}^{(-\frac{1}{2},-\frac{1}{2})}(z,\bar{z})&=&(P_{-}\fsH_{(n)}M_p)^{\al\be}e^{-\phi(z)/2}
S_{\al}(z)e^{i\frac{\alpha'}{2}p\cd X(z)}e^{-\phi(\bar{z})/2} S_{\be}(\bar{z})
e^{i\frac{\alpha'}{2}p\cd D \cd X(\bar{z})}\lam\otimes\sigma_3\sigma_1,\nonumber\\
V_{C}^{(-\frac{3}{2},-\frac{1}{2})}(z,\bar{z})&=&(P_{-}\fsC_{(n)}M_p)^{\al\be}e^{-3\phi(z)/2}
S_{\al}(z)e^{i\frac{\alpha'}{2}p\cd X(z)}e^{-\phi(\bar{z})/2} S_{\be}(\bar{z})
e^{i\frac{\alpha'}{2}p\cd D \cd X(\bar{z})}\lam\otimes\sigma_1
\label{d4Vs}
\eeqa

This provides a very interesting selection rule for all non-BPS brane amplitudes involving world volume fermion fields.

 \vskip 0.1in

 We highlight the crucial fact that the correlation function between two spin operators and one world sheet fermion field of non-BPS branes  ( even with some more fermion fields and/or currents coming from scalar /gauge in different pictures) is non-zero however due to postulating CP factors we realize that the amplitude of two fermion fields and an arbitrary numbers of gauge/scalar (even in different pictures) and one  tachyon makes no sense in the world volume of non-BPS branes.
 \vskip 0.1in


\vskip.2in

As an instance let us talk about the amplitude of two closed string  Ramond-Ramond and one tachyon. We claim this amplitude has zero value so there is no need to carry out all the correlation functions including four spin operators and one fermion field. Here is its argument.

\vskip.1in

We 	believe  that the correlation function between four spin operators and one world sheet fermion field of non-BPS branes  is non-zero, however, due to postulating CP factors for $C^{-1}C^{-1} T^0$ ($\Tr(\sigma_3\sigma_1\sigma_3\sigma_1\sigma_1)=0$) we realize that the amplitude of two closed string  Ramond-Ramond  ( even with  their different pictures with the same or different chirality) and one  tachyon makes no sense in the world volume of non-BPS branes.

\vskip.1in

The conclusion is that there is no coupling for any ordering of $CCT$ amplitude in the world volume of non-BPS branes, so there should not be any coupling between them  in field theory either.

\vskip 0.3in

\subsection{ The four point  $<V_{C} V_{T} V_{\bar\psi}V_{\psi} >$ amplitude in IIB }

\vskip 0.2in

 From now on we would like to compute the amplitude of one closed string Ramond-Ramond ( which moves in the bulk), two fermion fields and one tachyon in the world volume of non-BPS branes.

 \vskip.2in

 Motivation for carrying out this S-matrix is to actually obtain  all infinite higher derivative corrections of two fermions and two tachyons to all orders in $\alpha'$ and to fix precisely their coefficients as well. The extremely important point should be made is that, for sure these corrections can not be applied for brane anti brane systems (one might go through \cite{Hatefi:2012cp} for further details), thus these corrections which we are getting to derive are just related to two fermions and two tachyons of non-BPS branes and not to brane-anti brane systems.

 \vskip.1in
Some remarks  about our amplitude $CT\bar\psi \psi$ must be highlighted.
\vskip.2in

  The first fact  is that both two fermion fields here should have different chiralities which means that our calculation makes sense just for IIB super string theory. 
  
  This amplitude has zero result for IIA and the reason for this sharp conclusion is as follows:

\vskip.2in

  The multiplication of two spin operators with the same chirality gives us a vector (the same thing so happens for two spin operators of closed string RR). The multiplication of these two vectors with one  fermion field (the vertex of tachyon in zero picture consists of a vector in space-time) gives us three vectors and the multiplication of three vectors can not give rise a singlet.
   \vskip.1in
   Hence we immediately conclude that the correlation function of four spin operators (with the same chirality) and one fermion field has zero value. 
   \vskip.1in
   Therefore $<V_{C} V_{T} V_{\bar\psi}V_{\psi} >$ in IIA has zero result which means that neither there are   first order couplings between one RR,  two fermions and one tachyon in IIA nor infinite higher derivative corrections .  
   
   \vskip.1in
   Rather than the multiplication of two spin operators with different chirality (in IIB) gives us a singlet (identity matrix) and from RR we get a vector. Thus essentially the multiplication of a vector (tachyon in zero picture) and an identity matrix ( two spin operators with different chiralities) and a vector ( from RR) gives us a singlet. Therefore $<V_{C} V_{T} V_{\bar\psi}V_{\psi} >$
  just makes sense for IIB string theory.

\vskip.3in



Hence $<V_{C} V_{T} V_{\bar\psi}V_{\psi} >$ amplitude should be looked as follows

\begin{eqnarray}
{\cal A}^{  C T \bar\psi \psi } & \sim & \int dx_{1}dx_{2}dx_{3} dx_{4}dx_{5}
  \lan
V_{T}^{(0)}{(x_{1})}   V_{\bar\psi}^{(-1/2)}{(x_{2})}V_{\psi}^{(-1/2)}{(x_{3})} V_{C}^{(-1)}{(x_{4},x_{5})}
 \ran,\labell{sstring}\eeqa

\vskip 0.1in

We just need to take into account one ordering of the amplitude  so we choose it to be $\Tr(\lam_1\lam_2\lam_3)$ as usual. Having substituted our vertices in the S-matrix, some extraordinary tool is needed. 
\vskip 0.1in

Indeed one has to look for  the correlation function of one fermion field and four spin operators in ten dimensions of space time ( in order to have  non zero result in IIB string theory three spin operator have to have the same chirality and the fourth spin operator must carry different chirality). This correlator at one loop level has been derived   in \cite{Hartl:2010ks} but we are looking for it at tree level computations. Therefore in  order to find it out at tree level, one has to set all theta functions to an identity matrix. Another important point is in order.

\vskip.4in

In order to have the full result of the amplitude of $<V_{C} V_{T} V_{\bar\psi}V_{\psi} >$, the special gauge fixing ( namely gauge fixing on the location of open strings ) is necessary, in such a way that the needed correlator would be found as :

\beqa
 <\psi^a(x_1): S_\al(x_4) : S_\be(x_5): S_\ga(x_2): S^{\dot\delta}(x_3)>& =& \frac{(x_{45} x_{42} x_{52})^{-3/4}(I^a_{\al\be\ga\dot\delta})}{\sqrt{2}(x_{12}  x_{13}  x_{14}  x_{15})^{1/2}(x_{43}  x_{53}  x_{23})^{1/4}} \nonumber\eeqa

where

\beqa
I^a_{\al\be\ga\dot\delta}& =&\bigg[ \frac{C_\ga {}^{\dot\delta} }{x_{23}}  (\ga^a \, C)_{\al \be}  x_{13}  x_{42}  x_{52} +   \frac{C_\al {}^{\dot\delta} }{x_{43}}  (\ga^a \, C)_{\be \ga}  x_{13}  x_{45}   x_{42}  -  \frac{C_\be {}^{\dot\delta} }{x_{53}}   (\ga^a \, C)_{\al \ga}   x_{13}   x_{45}  x_{52} \nonumber\\&& -   \frac{1}{2}   (\ga^{\nu } \ga^{a} \, C)_\ga {}^{\dot\delta}   (\ga_\nu \, C)_{\al \be}   x_{14}   x_{52}
  +   \frac{1}{2} \; (\ga^{\nu } \ga^{a} \, C)_\al {}^{\dot\delta} \, (\ga_\nu \, C)_{\be \ga}  x_{12}  x_{45}
  \bigg]
\eeqa



\vskip 0.1in

Having applied above correlator inside the amplitude, we get to

\beqa {\cal A}^{C T \bar\psi \psi}&\sim& \int
 dx_{1}dx_{2}dx_{3}dx_{4} dx_{5}\,
(P_{-}\fsH_{(n)}M_p)^{\al\be} \bar u_1^{\ga} u_2^{\dot\delta}  (x_{23}x_{24}x_{25}x_{34}x_{35}x_{45})^{-1/4}\Tr(\sigma_1\sigma_3 I \sigma_3\sigma_1) I_{1}
\nonumber\\&&\times (\alpha' ik_{1a}) <\psi^a(x_1): S_\al(x_4) : S_\be(x_5): S_\ga(x_2): S^{\dot\delta}(x_3)>
 \Tr(\lam_1\lam_2\lam_3),\labell{125}\eeqa where
$x_{ij}=x_i-x_j,x_4=z=x+iy,x_5=\bar z=x-iy$, and
\beqa
I_{1} &=&{<:e^{\alpha' ik_1.X(x_1)}:e^{\alpha' ik_2.X(x_2)}
:e^{\alpha' ik_3.X(x_3)}:e^{i\frac{\alpha'}{2}p.X(x_4)}:e^{i\frac{\alpha'}{2}p.D.X(x_5)}:>}
\nonumber
\eeqa

Concerning Wick theorem we find
\beqa
I_{1}=   |x_{12}|^{\alpha'^2 k_1.k_2}|x_{13}|^{\alpha'^2 k_1.k_3}|x_{14}x_{15}|^{\frac{\alpha'^2}{2} k_1.p}|x_{23}|^{\alpha'^2 k_2.k_3}|
x_{24}x_{25}|^{\frac{\alpha'^2}{2} k_2.p}
|x_{34}x_{35}|^{\frac{\alpha'^2}{2} k_3.p}|x_{45}|^{\frac{\alpha'^2}{4}p.D.p}
\nonumber\eeqa

\vskip 0.2in

Setting the correlators in the integrand, we realize the fact that  the amplitude has the property of  SL(2,R) invariance.  Let us define the Mandelstam variables as

\
\beqa
s&=&\frac{-\alpha'}{2}(k_1+k_3)^2, \quad t=\frac{-\alpha'}{2}(k_1+k_2)^2, \quad u=\frac{-\alpha'}{2}(k_3+k_2)^2
\nonumber
\eeqa

\vskip 0.1in

Performing the gauge fixing on the location of open strings as  $(x_1=0,x_2=1,x_3=\infty)$, we  may find the final form of the amplitude  to be ready to take integrations on the location of closed string as follows

\beqa
{\cal A}^{C T \bar\psi \psi}=(P_{-}\fsH_{(n)}M_p)^{\al\be} (2\alpha' ik_{1a}) \bar u_1^{\ga} u_2^{\dot\delta}  \frac{1}{\sqrt{2}} \int
 dz d\bar z |z|^{2t+2s-1}|1-z|^{2t+2u-3/2} (z-\bar z)^{-2(t+s+u)-3/2}
\nonumber\\\times  \bigg[C_\ga {}^{\dot\delta} (\ga_a C)_{\al\be}|1- z|^2
-   C_\al^{\dot\delta}  (z-\bar z)(1-z)(\ga^a C)_{\be\ga}+ C_\be ^{\dot\delta} (\ga^a C)_{\al\ga}
(z-\bar z)(1-\bar z)\nonumber\\  -\frac{1}{2} (\ga^{\nu } \ga^{a} \, C)_\ga {}^{\dot\delta}   (\ga_\nu \, C)_{\al \be}   z (1-\bar z)
-  \frac{1}{2} \; (\ga^{\nu } \ga^{a} \, C)_\al {}^{\dot\delta} \, (\ga_\nu \, C)_{\be \ga}(z-\bar z)
 \bigg]\Tr(\lam_1\lam_2\lam_3)
\labell{amp3q},\eeqa

\vskip 0.2in

One important check of our amplitude and in particular our correlators is  indeed producing all the leading singularities  in which our computation  takes care of it. In fact by sending $x_1$ to $x_2$ we get to the correct correlator of four spin operators where two of them carry different chirality \cite{Hartl:2010ks} and this is one more test in favor of our computations.

\vskip 0.2in

The double integrals  should be performed  on the closed string position to actually get the entire result (further details should be found in  \cite{Fotopoulos:2001pt}, \cite{Hatefi:2012wj} ). Hence   the final result for our S-matrix is gotten as:

 \beqa
{\cal A}^{C T \bar\psi \psi}&=& \Tr(\lam_1\lam_2\lam_3) (P_{-}\fsH_{(n)}M_p)^{\al\beta}(2\alpha' ik_{1a}) \bar u_1^{\ga} u_2^{\dot\delta}
 \frac{1}{\sqrt(2)}
  \bigg[C_\ga ^{\dot\delta} (\ga_a C)_{\al\be}  L_1 +\bigg(- C_\al ^{\dot\delta}  (\ga^a C)_{\be\ga}\nonumber\\&&
  + C_\be ^{\dot\delta}(\ga^a C)_{\al\ga} \bigg )\frac{-s L_2}{(-s-u+\frac{1}{4})}
+\frac{1}{2} \bigg( C_\al ^{\dot\delta}  (\ga^a C)_{\be\ga}+ C_\be ^{\dot\delta}(\ga^a C)_{\al\ga} \bigg) L_3
+  \frac{1}{2}  (\ga^{\nu } \ga^{a} C)_\ga ^{\dot\delta}  (\ga_\nu  C)_{\al\be}
\nonumber\\&&\times\frac{(-u-\frac{1}{4})}{(-t-u-\frac{1}{4})}  L_{1}
+\bigg(-\frac{1}{2}(\ga^{\nu } \ga^{a} C)_\al ^{\dot\delta}  (\ga_\nu  C)_{\be \ga}- \frac{1}{4}  (\ga^{\nu } \ga^{a}  C)_\ga
^{\dot\delta}  (\ga_\nu C)_{\al\be}\bigg)
  L_2\bigg] \label{es23}\eeqa

with

\beqa
L_1&=&(2)^{-2(t+s+u)-3/2}\pi{\frac{\Gamma(-u-\frac{1}{4})
\Gamma(-s+\frac{1}{2})\Gamma(-t)\Gamma(-t-s-u-\frac{1}{4})}
{\Gamma(-u-t-\frac{1}{4})\Gamma(-t-s+\frac{1}{2})\Gamma(-s-u+\frac{1}{4})}},\nonumber\\
L_2&=&(2)^{-2(t+s+u)-1/2}\pi{\frac{\Gamma(-u+\frac{1}{4})
\Gamma(-s)\Gamma(-t+\frac{1}{2})\Gamma(-t-s-u+\frac{1}{4})}
{\Gamma(-u-t+\frac{3}{4})\Gamma(-t-s+\frac{1}{2})\Gamma(-s-u+\frac{1}{4})}},\nonumber\\
L_3&=&(2)^{-2(t+s+u)+1/2}\pi{\frac{\Gamma(-u+\frac{3}{4})
\Gamma(-s+\frac{1}{2})\Gamma(-t)\Gamma(-t-s-u+\frac{3}{4})}
{\Gamma(-u-t+\frac{3}{4})\Gamma(-t-s+\frac{1}{2})\Gamma(-s-u+\frac{5}{4})}},\nonumber\\
\label{Ls}
\eeqa


By postulating the internal degrees of freedom (CP factor) for all strings in the presence of non-BPS branes, we expect to get the same answer for our amplitude in the other different pictures. Hence we expect to get exactly the same result for the amplitude of $<V_C^{-1}(z,\bar z) V_T^{-1}(x_1) V_{\bar\Psi}^{1/2}(x_2)V_{\Psi}^{-1/2}(x_3)>$.

\vskip.2in

Note also that by removing tachyons we are set to BPS branes in which they do not carry CP factor thus the kinetic term of fermion fields (even in the presence of  non-BPS branes)  is just

\beqa
 2\pi\alpha' \Tr(\bar\Psi \ga^\mu D_\mu\Psi)\nonumber\eeqa 
 
 and in particular it does not carry internal degrees of freedom in field theory side. Therefore the kinetic term of fermion fields keeps fixed even in the world volume of  non-BPS branes in the DBI effective action.

\vskip.2in
 By taking a look at the final result of our amplitude, we reveal that it does have non-zero couplings for diverse cases. One good remark should be made as follows.
 \vskip.1in 
  
  Here we are not dealing with massless strings so the expansion as it stands, is not low energy expansion. How can we  expand the amplitude? 
  
   \vskip.1in
  The answer is that, the amplitude should be expanded such that all the singularities (massless/tachyonic poles) are indeed produced in comparison with the effective field theory arguments.

 \vskip .2in

 Let us mention some hints from string theory 's point of view. In the last section we proved that by making use of the internal degrees of freedom, neither there are couplings between two fermion fields and one tachyon nor between two tachyons and one fermion field. 
  \vskip .1in
 Having set this remark, we observe that there is no any kind of tachyon/fermion pole for this amplitude. Therefore we should not expect to have any $s,t$-channel pole in this amplitude.

  \vskip .1in

Hence,  by applying this remark to all the Gamma functions of our amplitude we might discover  that both $t,s$ must not send to $0,\frac{1}{2}$. We will talk about the other conditions for our amplitude later on.

\vskip .1in
 The other fact which must be highlighted is that by applying all CP factors we have checked that there is no any double poles including fermion-fermion, fermion-tachyon,  tachyon-scalar, fermion-gauge (scalar), tachyon-tachyon and gauge-gauge (scalar). 
   \vskip .2in
 
 For example one may talk about the coupling between two fermions and one scalar from left hand side and the coupling between two tachyons and one scalar from the middle and the coupling between one RR and one tachyon from right hand side. However this diagram does not have any contribution  given the fact that coupling between two tachyons and one scalar is zero (although the amplitude of two tachyons and one scalar does have non zero CP factor as shown by $\Tr(\sigma_3\sigma_2\sigma_1)=-2i$).

 \vskip .4in






\begin{center}
\begin{picture}
 (600,100)(0,0)
\Line(25,105)(75,70)\Text(50,105)[]{$\bar\Psi_{2}$}
\Line(25,35)(75,70)\Text(50,39)[]{$\Psi_{3}$}
\Line(125,70)(75,70)\Text(105,75)[]{$T$}
\Line(25,70)(75,70) \Text(50,75)[]{$T_1$}
\Gluon(125,70)(175,105){4.20}{5}\Text(145,105)[]{$C_{p}$}
\Line(295,105)(345,70)\Text(320,105)[]{$\bar\Psi_{2}$}
\Line(295,35)(345,70)\Text(310,35)[]{$\Psi_{3}$}
\Gluon(345,70)(395,105){4.20}{5} \Text(365,105)[]{$C_{p}$}
\Line(345,70)(395,35)\Text(365,52)[]{$T_1$}
\Vertex(345,70){1.5}
\Text(345,20)[]{(b)}

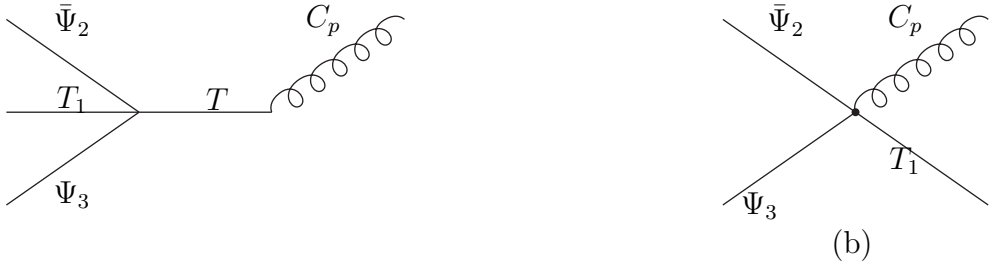
\end{picture}\\ {\sl Figure 1 : The Feynman diagrams corresponding to the four point function of $CT_1\bar\Psi_2 \Psi_3 $.}
\end{center}

\vskip .2in

 On the other hand (as it has been drawn) this amplitude should have infinite tachyon poles for    $n=p+1$  case, as there are non-zero couplings between two fermions and two tachyons  in the world volume of non-BPS branes ( the CP factor for $A^{\bar\psi^{-1/2} \psi^{-1/2} T^{-1}T^{0}}$  has also non zero value $ \Tr(\sigma_3I\sigma_2\sigma_1)=-2i $).  Also the amplitude does have infinite contact interactions which we are not going to explore them in this paper.

\vskip 0.3in

 Here we insist of performing direct computations to see whether or not the  universal conjecture for infinite higher derivative corrections of type II string theory works out (even for non-BPS fermionic amplitudes).

  \vskip .1in

  In order to  derive all  non-Abelian and the infinite higher derivative corrections of two tachyons and two fermions  (to all orders in $\alpha'$) and to produce all the infinite $u+s'+t'$ tachyon poles in string amplitude, one should have the expansion of the amplitude of two tachyons and two fermions and  should also know the final result of the amplitude of  ($A^{\bar\psi\psi T T}$)  to be able to  precisely produce
all the  coefficients of $a_{n,m},b_{n,m}$ in field theory side. 
\vskip 0.2in 
 
 Note that essentially we need to compare these coefficients at each order of $\alpha'$ with string coefficients in favor of obtaining all the infinite higher derivative corrections of two tachyons and two fermions in the world volume of non-BPS branes.

\vskip .2in

 Namely one has to have the complete form of both amplitudes. Thus all the following orderings must be taken into account :

\beqa
\Tr(\lam_1\lam_2\lam_3\lam_4),\Tr(\lam_1\lam_2\lam_4\lam_3),\Tr(\lam_1\lam_3\lam_2\lam_4)\nonumber\\ \Tr(\lam_1\lam_3\lam_4\lam_2),\Tr(\lam_1\lam_4\lam_3\lam_2),\Tr(\lam_1\lam_4\lam_2\lam_3),\nonumber
\eeqa

\vskip .2in

The other fact which must be regarded is that the final result of the open string amplitudes ( like $A^{\bar\psi^{-1/2} \psi^{-1/2} T^{-1}T^{0}}$) must respect all symmetries. Here the final answer should respect all symmetries related to two fermions and two tachyons amplitude, basically it has to be totally anti symmetric with respect to interchanging fermions, furthermore it should be symmetric under interchanging two tachyons . 

 \vskip .1in
In order to respect these symmetries,  one has to take into account all Chan-Paton factors and in particular consider all six possible orderings  of this amplitude (for four point open super strings).

\vskip.2in

 Now let us first carry out this amplitude for $\Tr(\lam_1\lam_2\lam_3\lam_4)$ ordering as:

\beqa
{\cal A}^{\bar \Psi,\Psi,T,T} & = & (-8 T_p 2^{1/2}) \int dx_1 dx_2  dx_3
 dx_{4}
 \Tr\lan
V_{\bar \Psi}^{(-1/2)}(x_1)V_{\Psi}^{(-1/2)}(x_2)V_{T}^{(-1)}(x_3)V_T^{0}(x_4)\ran
\nonumber\\&= & (-8 T_p )\int dx_1 dx_2
  dx_3  dx_{4}
|x_{12}|^{\alpha'^2 k_1\cdot k_2}|x_{13}|^{ \alpha'^2  k_1\cdot k_3}|x_{14}|^{ \alpha'^2   k_1\cdot k_4}|x_{23}|^{\alpha'^2 k_2\cdot k_3}|x_{24}|^{ \alpha'^2   k_2\cdot k_4} \nonumber\\&&\times  |x_{34}|^{\alpha'^2 k_3\cdot k_4} x_{12}^{-1}x_{13}^{-\frac{1}{2}}(x_{41}x_{42})^{-\frac{1}{2}}x_{23}^{-\frac{1}{2}}\Tr(\lam_1\lam_2\lam_3\lam_4)\Tr(\sigma_3 I\sigma_2\sigma_1)
( \alpha' ik_{4a})\bar u_1^A (\gamma^a )_{AB} u_2^B\nonumber\eeqa

\vskip.4in

 One can show that the amplitude is now   $SL(2,R)$ invariant. To get the final answer interms of Gamma function one has to carry out gauge fixing as $(x_1,x_2,x_3,x_4)=(0,x_2,1,\infty)$  where at the end we get to

\beqa
{\cal A}^{\bar \Psi,\Psi,T,T}  &= & (-8 T_p)\Tr(\sigma_3 I\sigma_2\sigma_1) \bar u_1^A(\ga^{a})_{AB}u_2^B (\alpha'ik_{4a})\int_0^{1} dx_2 x_2^{-2t-1}(1-x_2)^{-2u-1} \Tr(\lam_1\lam_2\lam_3\lam_4)
\nonumber\\ & =&(-8 T_p) \bar u_1^A(\ga^{a})_{AB}u_2^B (\alpha'ik_{4a}) \Tr(\lam_1\lam_2\lam_3\lam_4)\Tr(\sigma_3 I\sigma_2\sigma_1)\frac{\Gamma(-2t)\Gamma(-2u)}{\Gamma(-2t-2u)}\eeqa

\vskip.1in

 However one has to find out all the other 5 possible orderings as well. For instance for   $\Tr(\lam_1\lam_3\lam_2\lam_4)$ the corrected gauge fixing is $(x_1,x_3,x_2,x_4)=(0,x_3,1,\infty)$ also note that in all possible orderings the picture of open strings should be kept fixed.

 \vskip.2in

 By extracting the CP factors the final and complete form of the amplitude is


\beqa
{\cal A}^{\bar \Psi,\Psi,T,T} & = &(-8 T_p ) \bar u_1^A(\ga^{a}_{AB})u_2^B (\alpha'ik_{4a})\Tr(\sigma_3 I\sigma_2\sigma_1)\bigg(l_1\frac{\Gamma(-2t)\Gamma(-2u)}{\Gamma(-2t-2u)}-l_2 \frac{\Gamma(-2t)\Gamma(-2s)}{\Gamma(-2t-2s)}\nonumber\\&&
+il_3\frac{\Gamma(-2u)\Gamma(-2s)}{\Gamma(-2u-2s)}\bigg)\labell{amp1}
\eeqa
where $l_1,l_2,l_3$ are defined as
\beqa
l_1=\frac{1}{2}\bigg(\Tr(\lam_1\lam_2\lam_3\lam_4)+\Tr(\lam_1\lam_4\lam_3\lam_2)\bigg)\nonumber\\
l_2=\frac{1}{2}\bigg(\Tr(\lam_1\lam_2\lam_4\lam_3)+\Tr(\lam_1\lam_3\lam_4\lam_2)\bigg)\nonumber\\
l_3=\frac{1}{2}\bigg(\Tr(\lam_1\lam_3\lam_2\lam_4)+\Tr(\lam_1\lam_4\lam_2\lam_3)\bigg)\nonumber
\eeqa

The very non trivial question would be related to the expansion of the amplitude, basically  how to expand the amplitude such that all massless poles are not removed.
 Making use of the on-shell condition  
 
 \beqa
  s+t+u=-\frac{1}{2}\nonumber\eeqa  we might realize that all Mandelstam variables should not be sent to zero. Remember there is a non zero coupling between two fermions and one gauge field, moreover there is a non zero  coupling  between two tachyons and one gauge field. Thus we realize that there must be a massless t-channel pole for this amplitude.

 \vskip.2in

 Also note that the amplitude must be symmetric with respect to interchanging  $s$- and $u$ (under interchanging  both tachyons) so the correct expansion which satisfies on-shell condition is  indeed

\beqa
t\rightarrow 0,\qquad s,u\rightarrow -\frac{1}{4}\label{143e}
\eeqa
Let us write down the above momentum expansion just in terms of momenta,(see \cite{Hatefi:2012wj,Hatefi:2010ik})
\beqa
(k_1+k_2)^2,k_1\inn k_3,k_2\inn k_3\rightarrow 0 \nonumber
\eeqa

The above relation is the momentum expansion that we have looked for. Using  $u'=u+1/4=-\alpha'k_2\inn k_3$ and $s'=s+1/4=-\alpha'k_1\inn k_3$ we can derive new on-shell relation as $s'+t+u'=0$.

\vskip.2in

One can make use of the new relation ($s'+t+u'=0$) and also use the above change of variables to actually derive the final form  of the amplitude as follows:

\beqa
{\cal A}^{\bar \Psi_1,\Psi_2,T_3,T_4} & \!=\! &(-8 T_p )\bar u_1^A(\ga^{a}_{AB})u_2^B (\alpha'ik_{4a})\Tr(\sigma_3 I\sigma_2\sigma_1)
\bigg(l_1\frac{\Gamma(2u'+2s')\Gamma(\frac{1}{2}-2u')}{\Gamma(\frac{1}{2}+2s')}\nonumber\\&&-l_2 \frac{\Gamma(2u'+2s')\Gamma(\frac{1}{2}-2s')}{\Gamma(\frac{1}{2}+2u')}+il_3\frac{\Gamma(\frac{1}{2}-2s')\Gamma(\frac{1}{2}-2u')}{\Gamma(1-2s'-2u')}
\bigg)\labell{amp4}
\eeqa

If we expand our amplitude around  \reef{143e} we eventually get

\beqa
{\cal A}^{\bar \Psi_1,\Psi_2,T_3,T_4} & = &-8 T_p
  \bar u_1^A(\ga^{a}_{AB})u_2^B (\alpha'ik_{4a})\Tr(\sigma_3 I\sigma_2\sigma_1)    \labell{amp52}\\&&
  \bigg(\frac{l_1-l_2}{-2t}+\sum_{n,m=0}^{\infty}\bigg[a_{n,m}(l_1 u'^ns'^m-l_2 s'^n u'^m)+il_3 b_{n,m}(s'^nu'^m+ s'^mu'^n)\bigg]\bigg)\nonumber
\eeqa

\vskip.2in

 Ultimately after using the expansion in \reef{143e} we are able to derive all the correct  coefficients of $a_{n,m},b_{n,m}$ as follows
\beqa
&&a_{0,0}=2\ln(2),b_{0,0}=\frac{\pi}{2},a_{1,1}=-12\z(3)+\frac{32}{3}\ln(2)^3+\frac{4\pi^2}{3}\ln(2),\,b_{1,1}=\frac{\pi}{3}(-\pi^2+24\ln(2)^2),\nonumber\\
&&a_{1,0}=\frac{2\pi^2}{3}+4\ln(2)^2,\,b_{1,0}=2\pi\ln(2),\,a_{0,1}=-\frac{\pi^2}{3}+4\ln(2)^2,\nonumber\\
&&a_{2,0}=8\z(3)+\frac{16}{3}\ln(2)^3+\frac{8\pi^2}{3}\ln(2),b_{2,0}=\frac{\pi}{3}(\pi^2+12\ln(2)^2)\nonumber\\
&&\,a_{0,2}=8\z(3)+\frac{16}{3}\ln(2)^3-\frac{4\pi^2}{3}\ln(2),\,b_{1,2}=\frac{4\pi}{3}(12\ln(2)^3-3\z(3)),\nonumber\\
&&b_{3,0}=\frac{4\pi}{3}(\pi^2\ln(2)+4\ln(2)^3+6\z(3)),\cdots\label{coefs}
\eeqa

Note that $b_{n,m}$
's coefficients are symmetric. The massless gauge t-channel pole can just be resulted by taking into account all the kinetic terms of fermions, gauges and tachyons.

\beqa
-T_p \frac{2\pi\alpha'}{2}
  \Tr\left( \bar \Psi\gamma ^a D_a\Psi + D_aTD^aT-\frac{(2\pi\alpha')}{2}F_{ab}F^{ba}\right)\labell{kinetic terms}
\eeqa

where
\beqa
D_a\Psi &=&\partial_a\Psi-i[A^a,\Psi], D_aT=\partial_a T-i[A^a,T],D_i\Psi =\partial_i\Psi-i[\phi^i,\Psi], \nonumber\eeqa

In the other words the massless t-channel gauge pole should be produced by considering the following rule
\beqa
V^{a,\alpha}(\bar \Psi_1,\Psi_2, A)G^{ab,\alpha\beta}(A)V^{b,\beta}(A,T_3,T_4)\nonumber
\eeqa

 \begin{center}
\begin{picture}
 (600,100)(0,0)
\Line(25,105)(75,70)\Text(50,105)[]{$\bar\Psi_{1}$}
\Line(25,35)(75,70)\Text(50,39)[]{$\Psi_{2}$}
\Photon(125,70)(75,70){4.20}{5}\Text(105,90)[]{$A$}
\Line(125,70)(175,35) \Text(150,46)[]{$T_3$}
\Line(125,70)(175,105) \Text(145,105)[]{$T_4$}
\Line(295,105)(345,70)\Text(320,105)[]{$\Psi_2$}
\Line(295,35)(345,70)\Text(310,35)[]{$\bar\Psi_{1}$}
\Line(345,70)(395,105) \Text(365,105)[]{$T_3$}
\Line(345,70)(395,35) \Text(370,46)[]{$T_4$}
\Vertex(345,70){1.5}
\Text(345,20)[]{(b)}

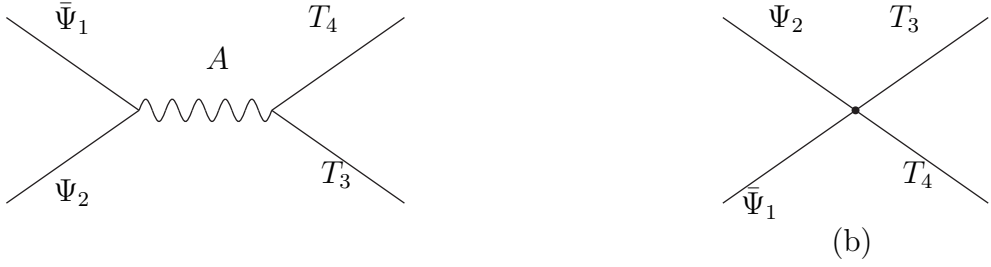
\end{picture}\\ {\sl Figure 2 : The Feynman diagrams corresponding to the amplitude of $  \bar\Psi_1 \Psi_2 T_3T_4 $.}
\end{center}

 such that
 \beqa
  G^{ab,\alpha\beta}(A)&=&\frac{i\delta^{ab}\delta^{\alpha\beta}}{(2\pi\alpha')^2T_p t}\nonumber\\
 V^{a,\alpha}(\bar\Psi_1,\Psi_2, A)&=&T_p(2\pi\alpha')\bar u_1^A\ga^a_{AB}u_2^B\left(\Tr(\lam_1\lam_2\lam^\alpha)-\Tr(\lam_2\lam_1\lam^\alpha)\right)\nonumber\\
 V^{b,\beta}(A,T_3,T_4)&=&T_p(2\pi i\alpha')(k_3^b-k_4^b)\left(\Tr(\lam_4\lam_3\lam^\beta)-\Tr(\lam_3\lam_4\lam^\beta)\right)
 \label{massless tchannel}
 \eeqa

\vskip.2in

The other question which should be addressed is how to produce the contact terms.
Let us write down the zeroth order of contact terms of the amplitude as
\beqa
{\cal A}^{\bar \Psi_1,\Psi_2,T_3,T_4} & = &-8 T_p (2i)
  \bar u_1^A(\ga^{a}_{AB})u_2^B (\alpha'ik_{4a}) \left[a_{0,0}(l_1 -l_2 )+2il_3 b_{0,0}\right]\nonumber
\eeqa

The above contact interactions can be derived by writing down some suitable couplings.

\vskip.2in

It is worth  emphasizing that these new couplings must have the same  order like the kinetic interaction of the fermion fields. The momentum conservation along the world volume of brane $(k_1+k_2+k_3+k_4)^a=0 $ as well as  on shell conditions for fermion fields $\gamma^a k_{1a} \bar u^A=\gamma^a k_{2a}  u^B=0 $ should have been applied. Hence contact interaction at first order can be summarized as

\beqa
\alpha'T_p\Tr\left(\frac{}{}a_{0,0}\left(\bar \Psi\ga^a\Psi TD_a T-\bar\Psi\ga^a\Psi D_aT T\right)+ib_{0,0}\left(\bar\Psi\ga^a T\Psi D_a T-\bar\Psi\ga^aD_aT\Psi T\right)\right)\labell{nonabcoup}
\eeqa

 It is not a big deal to show that
for Abelian groups and making use of just $(a_{0,1}-a_{1,0})$ term, the first non zero term
in \reef{amp52} can be precisely produced by expanding the effective action that appeared in page two and taking the following coupling
( for the expansion see \cite{Hatefi:2010ik}).

\beqa
2\pi^2\alpha'^2 T_p\bar \Psi\gamma^b \partial^a\Psi\partial_bT\partial_a T
\nonumber\eeqa

\vskip.2in

Now we come to one of our main goals  which is producing non-Abelian interactions of the scattering amplitude of one RR, one tachyon and two world volume fermion fields.
\vskip.2in

The coefficient of $
  \bar u_1^A(\ga^{a}_{AB})u_2^B ( i\alpha' k_{1a}) $ is the over all coefficient for  all terms inside the final result of the amplitude of $CT\bar\psi \psi$. According to the prescription which has been given in
\cite{Hatefi:2012rx,Hatefi:2012wj}, we expect to actually have explored all the non-Abelian couplings of two fermions and two tachyons in the world volume of non-BPS branes as :

 \vskip.1in

 \beqa
{\cal L}^{n,m}&=&\frac{\alpha'}{2}\alpha'^{n+m}T_p\left(a_{n,m}\Tr\left[\frac{}{}\cD_{nm}\left(\bar\Psi\ga^a\Psi T D_a T\right)+\cD_{nm}\left(T D_a T\bar\Psi\ga^a\Psi \right)\right.\right.\nonumber\\
&&\left.\left.-\frac{}{}\cD_{nm}\left(\bar\Psi\ga^a\Psi D_aT  T\right)-\cD_{nm}\left(D_aT  T\bar\Psi\ga^a\Psi \right)+h.c\right]\right.\nonumber\\
&&\left.+ib_{n,m}\Tr\left[\cD'_{nm}\left(\bar\Psi\ga^a T\Psi D_a T\right)+\cD'_{nm}\left(\Psi D_aT\bar\Psi \ga^a  T\right)\right.\right.\nonumber\\
&&\left.\left.-\cD'_{nm}\left(\bar\Psi\ga^a D_aT\Psi  T\right)-\cD'_{nm}\left(\Psi T\bar\Psi \ga^a  D_aT\right)+h.c\right]\frac{}{}\right)\labell{Lnm}
\eeqa

\vskip.2in

in such a way that the higher derivative operators $\cD_{nm}$ and $\cD'_{nm}$ are used to be defined 

\beqa
\cD_{nm}(EFGH)&\equiv&D_{b_1}\cdots D_{b_m}D_{a_1}\cdots D_{a_n}E  F D^{a_1}\cdots D^{a_n}GD^{b_1}\cdots D^{b_m}H\nonumber\\
\cD'_{nm}(EFGH)&\equiv&D_{b_1}\cdots D_{b_m}D_{a_1}\cdots D_{a_n}E   D^{a_1}\cdots D^{a_n}F G D^{b_1}\cdots D^{b_m}H\nonumber
\eeqa

\vskip.2in

 In fact the above couplings are non-Abelian extensions at the order of $(\alpha')^{1+n+m}$ of  $a_{0,0}$ and $b_{0,0}$ . One highly important remark about our notation is in order.

\vskip.2in

In the  term $\cD'_{nm}\left(\Psi D_aT\bar\Psi \ga^a  T\right)$ in \reef{Lnm}, first we need to apply all the higher derivative terms on it and then 
one should  rewrite it as $\left(\bar\Psi \ga^a  T\Psi D_aT\right)$.

\vskip.1in

This is another check in favor of universality conjecture for  higher derivative corrections  to all orders in $\alpha'$ for all non-BPS and BPS branes  \cite{Hatefi:2012rx}.


  \subsection{An infinite number of tachyon poles of $CT\bar\Psi\Psi$ for $p+1=n$ case}

\vskip.2in

  In order to show that we have derived all the infinite non-Abelian couplings of two fermions and two tachyons in the world volume of non-BPS branes \reef{Lnm}, in this section we are going to use those couplings to actually produce all the infinite tachyon poles of string amplitude.

\vskip.2in

Extracting the traces, we  just write down  all the singularities of the string amplitude  as

   \beqa
{\cal A}^{CT\bar\Psi\Psi}&=&\frac{32i \pi^2\mu'_p\beta'}{4(p+1)!}  \bar u_1^A(\ga^{a}_{AB})u_2^B
 \Tr(\lam_1\lam_2\lam_3)
\eps^{a_{0}\cdots a_{p}}H_{a_{0}\cdots a_{p}}\nonumber\\&&\times
(-2i\alpha' k_{1a})\sum_{n,m=0}^{\infty}\frac{e_{n,m}[s'^nt'^m -t'^ns'^m]}{(-t'-s'-u)}\label{nn89}\eeqa

  where we have normalized the string amplitude by $\frac{\beta'\mu'_p}{(2\pi)^{1/2}}$ and $\mu'_p$ is the RR charge of branes and $\beta'$ is the normalization of Wess-Zumino actions for non-BPS branes (further details can be understood in \cite{Hatefi:2012wj}).

\vskip.1in
 Note that we have  expanded the amplitude of $CT\bar\Psi\Psi$ out   such that  both conditions

 \beqa
 t+s+u=-p^ap_a-\frac{1}{4} \nonumber\eeqa

  which is momentum conservation along the world volume of brane
   and  the condition comes from Gamma function $ (-t-s-u-\frac{1}{4}=0)$   could
simultaneously be satisfied.

\vskip.2in

It is argued in \cite{Hatefi:2012wj,Hatefi:2012cp}
 that for non-BPS branes and  brane -anti brane systems the quantity
 $p^ap_a$ must tend to $\frac{1}{4},0$ accordingly. Also note that from our S-matrix it is kind of obvious that the expansion of fermionic amplitudes in the presence of tachyons is certainly different from the expansion of the amplitudes including RR, gauge/scalar in the presence of tachyons.
 
 \vskip.2in
 
The reason can be understood by taking a look at the concluded operator product expansions from   super ghost charges in such
 a way that after considering all symmetries here we do have $\Gamma(-t-s-u-\frac{1}{4})$ in the amplitude rather
than having usual $\Gamma(-t-s-u-\frac{1}{2})$ for bosonic amplitudes.

\vskip.2in

   In order to derive all the infinite tachyon poles of the string amplitude one must use the following Feynman rule :

  \beqa
{\cal A}&=&V^{\alpha}(C_{p},T)G^{\alpha\beta}(T)V^{\beta}(T,T_1,\bar\Psi_2,\Psi_3)\labell{amp5441}\eeqa
such that

\beqa
G^{\alpha\beta}(T) &=&\frac{i\delta^{\alpha\beta}}{(2\pi\alpha') T_p
(-k^2-m^2)}\nonumber\\
V^{\alpha}(C_{p},T)&=&2i\mu'_p\beta'(2\pi\alpha')\frac{1}{(p+1)!}\epsilon^{a_0\cdots a_{p}}H_{a_0\cdots a_{p}}\Tr(\Lambda^{\alpha})
\labell{Fey1}
\eeqa

\vskip.1in

 Notice that the propagator can be written as  $\frac{i\delta^{\alpha\beta}}{(2\pi\alpha') T_p
(-s'-t'-u)}$ and $\Tr(\Lambda^{\alpha})$ is just non-zero for Abelian  $\Lambda^{\alpha}$.

\vskip.2in

 Now
 the  $ V^{\beta}(T,T_1,\bar\Psi_2,\Psi_3)$  should be looked for by making use of the derived higher derivative non-abelian couplings of \reef{Lnm}.

\vskip.2in

  To obtain  two on-shell fermions, one on-shell tachyon and one off-shell tachyon's vertex operator, we need to work out with two different orderings of open strings as follows:

\beqa
\Tr(\lambda_2\lambda_3\lambda_1\lambda_{\beta}), \quad\quad\quad \Tr(\lambda_2\lambda_3\lambda_{\beta}\lambda_1)
\eeqa
where $\beta$ is related to the Abelian tachyon. Let us focus firstly  on $a_{n,m}$ such that by taking the following term   $\Tr\left(a_{n,m} \bar\Psi\ga^a\Psi T D_a T\right)$, we are able to derive the following vertices
\beqa
a_{n,m}(k.k_2)^m(k_1.k_2)^n \bar u\ga^a u (-ik_{4a})\nonumber\\
+a_{n,m}(k.k_2)^n(k_1.k_2)^m \bar u\ga^a u (-ik_{1a})
\eeqa

 where $k$ becomes  off-shell tachyon's momentum. Now we need to take into account the hermitian conjugate of this term as well as carrying out the same procedure to all the other terms in \reef{Lnm}. Adding all the terms up, we can derive the following vertex (just for the terms including the coefficient of $a_{n,m}$)

 \beqa
&&a_{n,m} \bar u\ga^a u \bigg(ik_{4a}-ik_{1a}\bigg)\bigg\{\bigg[-(k.k_2)^m(k_2.k_1)^n+(k_1.k_3)^n(k.k_3)^m-(k_3.k_1)^m(k_2.k_1)^n
\nonumber\\
&&+(k_3.k_1)^n(k_1.k_2)^m\bigg]
 -\bigg[-(k_2.k_1)^m(k_2.k)^n+(k.k_3)^n(k_1.k_3)^m-(k.k_2)^n(k_3.k)^m\nonumber\\
 &&+(k_3.k)^n(k_2.k)^m\bigg]\bigg\}
\nonumber\eeqa

\vskip.2in

  By using the momentum conservation along the world volume direction, applying equations of motion for fermion fields $\ga^a k_{2a} \bar u^A=\ga^a k_{3a} u^B=0$ and also by working out with the terms having  $b_{n,m}$ coefficients (as well as their conjugates )  we get to the final answer  for $ V_{\beta}^{b}(\bar\Psi_1,\Psi_2,T_3,T)$ as

  \beqa
   V_{\beta}^{b}(\bar\Psi_1,\Psi_2,T_3,T)&=& iT_{p} (\alpha')^{n+m+1}(a_{n,m}+ib_{n,m})
\bar u_1^{A}(\ga^a)_{AB} u_2^{B}  (-\alpha'ik_{1a})
\bigg\{\bigg[-(k.k_2)^m(k_2.k_1)^n\nonumber\\&&
+(k_1.k_3)^n(k.k_3)^m-(k_3.k_1)^m(k_2.k_1)^n+(k_3.k_1)^n(k_1.k_2)^m
 +(k_2.k_1)^m(k_2.k)^n\nonumber\\&&
 -(k.k_3)^n(k_1.k_3)^m+(k.k_2)^n(k_3.k)^m-(k_3.k)^n(k_2.k)^m\bigg]\bigg\}
  \Tr(\lam_1\lam_2\lam_3\lambda_{\beta})
 \labell{verpt}\eeqa

If we replace \reef{verpt}   into \reef{amp5441}, one can explore all the infinite  tachyon poles of  amplitude in field theory  as follows :

  \beqa
32i\beta'\mu'_p\frac{\eps^{a_{0}\cdots a_{p}}(-ik_{1a})\bar u_1^{A} (\ga^a)_{AB} u_2^B
H_{a_0\cdots a_{p}}}{(p+1)!(s'+t'+u)}\Tr(\lam_1\lam_2\lam_3)
\sum_{n,m=0}^{\infty} (a_{n,m}+ib_{n,m})[s'^{m}t'^{n}-s'^{n}t'^{m}]
\label{fields}\eeqa

\vskip 0.2in

  Now we would like to explicitly show  that the higher derivative terms in \reef{Lnm} are exact. In order to do so,  we first remove all the common factors from both string and field theory sides and do check both sides at each order of $\alpha'$.

\vskip 0.2in

 At  $\alpha'$ order we get the following coefficient in field theory side :
\beqa
 (a_{1,0}-a_{0,1})(s'-t')+(b_{1,0}-b_{0,1})(s'-t')&=&\pi^2(s'-t') \nonumber\eeqa

 where we have used the facts that $b_{n,m}$'s are symmetric  and in particular the coefficients in \reef{coefs} have been used. This coefficient is precisely equal to the coefficient in string amplitude as

 \beqa
 \pi^2  ( e_{1,0}-e_{0,1})(s'-t') \nonumber\eeqa

  At  $\alpha'^2$ order , we get
\beqa
&& (a_{1,1}+ib_{1,1})(s't'-s't')+i(b_{2,0}-b_{0,2})(s'^2-t'^2)+(a_{2,0}-a_{0,2})[s'^2-t'^2]\nonumber\\
&&=4\pi^2 ln(2)(s'^2-t'^2)
\nonumber\eeqa

which is equivalent to $\pi^2  ( e_{2,0}-e_{0,2})(s'^2-t'^2)$ in string amplitude.

\vskip 0.2in

Therefore all order $\alpha'$ checks can be made  to investigate that not only do we have exactly produced all the higher derivative corrections of two tachyons and two fermions but also all the infinite tachyon poles of  $<V_CV_TV_{\bar\Psi} V_{\Psi}>$ are precisely resulted.

\vskip 0.2in

 Thus we come to the fact that  these higher derivative couplings of two tachyons and two fermions in \reef{Lnm} are derived in a precise manner. It is worth to mention that these non-BPS corrections can not be used for brane anti brane systems as it is shown in \cite{Hatefi:2012cp}.

\vskip 0.2in
The final remarks are crucially in order. 

\vskip 0.2in

Indeed  by using $a_{0,0}=2\ln(2)$ and $b_{0,0}=\pi/2$, we understand the important fact that  the non-Abelian couplings at the order of  $\alpha'$ could not be matched by applying symmetric trace of non-Abelian couplings of \reef{Lnm}.

\vskip 0.2in

As it has been argued the above higher derivative corrections can not be used for brane anti brane systems. It would be interesting to actually find out these corrections for brane anti brane systems and also to see whether or not non-Abelian symmetric trace does work at order of $\alpha'$ for brane anti brane systems. 
\vskip 0.2in
The other open question would be related to solving  the ambiguity between ordinary trace and symmetrized trace of the amplitude of one RR and four tachyons in the world volume of brane anti brane systems which  is addressed in \cite{Hatefi:2012wj}.
\vskip 0.2in

Finally we refer to  super symmetric  action for having included more details and just point out that  the complete form of symmetrized action is still unknown \cite{Cederwall:1996pv,Cederwall:1996ri,Bergshoeff:1996tu}. We hope to address these issues in near future.

\vskip 0.2in

 \section{Conclusions}
 
In this paper we have considered the computation of a disk level four point correlation function, including one closed string Ramond-Ramond field in the bulk, one tachyon and two fermion fields ($<V_C V_T V_{\bar\Psi} V_\Psi>$) in the world volume of non-BPS branes. The aim of this paper was to find out the infinite higher derivative corrections of two fermions and two tachyons to all orders of $\alpha'$ in type IIB super string theory which are derived in \reef{Lnm}. 
\vskip.2in

Since there is no correction to the coupling of one Ramond-Ramond and one tachyon as follows

\beqa
\nonumber
2i\beta'\mu'_p(2\pi\alpha')\int_{\Sigma_{p+1}} C_p\wedge DT
\eeqa
we have understood that all the infinite tachyon poles must be produced by all the derived corrections in \reef{Lnm}, namely
we have checked these corrections by producing all the 
infinite $t'+s'+u$-channel tachyon poles  of the string amplitude for $p=n+1$ case.
Basically the presence of possible couplings between these fields in the Wess-Zumino action for non-BPS branes has been  investigated. We have also seen that due to internal degrees of freedom of open strings neither there are  fermion nor tachyon poles in the amplitude of $CT\bar \psi\psi$.
\vskip.2in

Due to Chan-Paton factors we also argued that there is no coupling between two closed string Ramond-Ramond field and one tachyon in the world volume of non-BPS branes. Therefore this amplitude does not make sense in type II super string theory.

 \section*{Acknowledgments}

 I would like to thank E.Witten, J. Polchinski, W.Lerche, A.Sen, K.S.Narain, N.Lambert, N.Arkani-Hamed, A.Sagnotti , G.Veneziano and C.Vafa for  useful discussions. I also acknowledge I.Antoniadis and L.Alvarez-Gaume and theory division at CERN for their hospitality where some part of this work was  carried out there.


\end{document}